\documentclass[aps,pra,twocolumn,superscriptaddress,floatfix]{revtex4-1}
\usepackage{amsmath}
\usepackage{amssymb}
\usepackage{epsfig}
\usepackage[colorlinks=true,citecolor=blue]{hyperref}
\usepackage{bbm}
\usepackage{newtxtext,newtxmath}

\graphicspath{{.}{./figs/}}

\newcommand{\ket}[1]{{\vert #1 \rangle}}
\newcommand{\ketbra}[2]{{\vert #1 \rangle\langle #2\vert}}
\newcommand{\proj}[1]{{\vert #1 \rangle\langle #1\vert}}
\newcommand{\braket}[2]{\langle{#1}|{#2}\rangle}
\newcommand* {\bra}[1]{\ensuremath{\langle {#1} |}}

\begin{document}

\title{Atomic Bell measurement via two-photon interactions}

\author{Carlos A. Gonz\'alez-Guti\'errez}
\affiliation{Instituto de Ciencias F\'isicas, Universidad Nacional Aut\'onoma de M\'exico, Apartado Postal 48-3, Cuernavaca, Morelos 62251, M\'exico}
\author{Juan Mauricio Torres} 
\affiliation{Instituto de F\'isica, Benem\'erita Universidad Aut\'onoma de Puebla, Apartado Postal J-48, Puebla 72570, M\'exico}

\date{\today}
\begin{abstract}
 We introduce a complete Bell measurement on atomic qubits based on two photon interactions with
 optical cavities and discrimination of coherent states of light. The dynamical system is described
 by the Dicke model for two three-level atoms interacting in two-photon resonance with a single-mode of the radiation field, which is known to effectively generate a non-linear two-photon interaction between the field and two states of each atom. For initial coherent states with large mean photon number, the field state is well represented by two coherent states at half revival time. 
For certain product states of the atoms, we prove the coherent generation of GHZ states with two atomic qubits and two orthogonal Schr\"odinger cat states as a third qubit. For arbitrary atomic states, we show that discriminating the two states of the field corresponds to different operations in the Bell basis of the atoms. 
 By repeating this process with a second cavity  prepared in  a phase-shifted coherent state, we demonstrate the  implementation of a complete Bell measurement. 
 Experimental feasibility of our protocols is discussed for cavity-QED, circuit-QED 
 and trapped ions setups.\end{abstract}
\maketitle
\section{Introduction}
\label{intro}
Bell measurements are crucial to implement quantum information
protocols such as quantum teleportation, superdense coding, and 
entanglement swapping \cite{Nielsen,Zeilinger}. These protocols play a key role in the 
nodes of a quantum repeater and to establish long-distance communication in a quantum network 
\cite{VanLoock06,Pirandola}. 
In a complete Bell measurement a two-qubit  system is probabilistically projected onto 
one of the four Bell states. 
For photonic qubits, it is possible to identify two of the four Bell states,
i.e. a 50\%-efficient Bell measurement,  using interference effects with linear optics \cite{Michler96,Suominen,Calsamiglia}.  The capability to surpass  this limit relies 
either on more resourceful techniques \cite{Grice2011,VanLoock} 
or on higher order optical interactions \cite{Bennett} that report low efficiency in the experiment \cite{Kim}.
In the case of atomic qubits, most experimental realizations of quantum teleportation
consider the implementation of a complete Bell measurement through entangling gates, such as a 
controlled-NOT  
or controlled-phase gate, that together with single qubit gates can map Bell sates onto product states in the computational basis \cite{Blatt,Barrett,Riebe}. 
A problem with this approach, however, is that high fidelity two-qubit 
gates are still experimentally difficult to achieve \cite{Schmidt-Kaler,Schmidt-Kaler2,Nolleke}. 

Motivated by the hybrid quantum repeater \cite{VanLoock06} that employs material
qubits and multiphoton coherent signals, a recently proposed alternative 
is to explore atom-photon interaction models  that directly generate Bell states of 
the atoms correlated with states of the field.
Using this approach, it was shown that unambiguous Bell state projections
can be implemented within the framework of the two-atom 
Tavis-Cummings model \cite{Torres2014,Torres2016,Bernad17}. 
There, the state of the atoms is postselected by 
projecting the states of the field onto nearly orthogonal coherent states. 
The great benefit is that the atomic states are not directly measured and their
projection occurs as postselection of the measured field. 
An imperfect efficiency 
relies on the fact that initial coherent states of the field in the Tavis-Cummings model 
do not evolve coherently during the interaction \cite{Jarvis,Torres2014}. This is clearly manifested in the non-perfect 
revivals of Rabi oscillations of atomic observables, similar to the well known
collapse and revival phenomena in the Jaynes-Cummings model \cite{Eberly}. 
The natural question that arises is whether
an atom-field model presenting perfect revivals of Rabi oscillations could better assist in the 
postselection of atomic Bell states. The answer to this question turns out to be positive as we
shall demonstrate.

In this paper we propose a complete atomic Bell measurement based on the two-photon two-atom
Dicke model in the rotating wave approximation \cite{Klimov} 
that presents nearly perfect revivals of Rabi oscillations. 
Similar to  previous work \cite{Torres2014,Torres2016}, the states of the qubits are encoded in 
a pair of two-level 
atoms that interact resonantly and sequentially with the field inside two optical cavities and the
atomic state is postselected by measuring the optical field.
The considered two-photon atom-field interaction model was first introduced as a generalization 
of the Jaynes-Cummings model \cite{Buck,Zubairy,Toor}, and later extended for multiatomic systems \cite{Jex,Klimov}. 
It has been proposed theoretically, but its experimental feasibility
has been analyzed in well controllable quantum optical systems \cite{Haroche}. 
Although we focus on a cavity QED implementation, 
two-photon or two-phonon interactions have 
also been studied and proposed in  circuit QED and trapped ions \cite{Plenio,Solano}, thus
making our proposal attractive to other architectures involving matter-field interaction.

The  paper is organized as follows. In Sec.~\ref{model}, we review the effective two-photon Dicke 
model as a limiting case of a general Dicke Hamiltonian of two three-level atoms and discuss the 
regime of validity in terms of the parameters of the system.
In Sec.~\ref{solution}, an approximate exact solution of the effective model in terms
of coherent states of the field is presented and its validity is
verified by comparing the fidelity with respect to exact numerical calculations.
Collapse and revival of Rabi oscillations  are studied in Sec.~\ref{Rabi}. 
The generation of tripartite entangled states is discussed in Sec.~\ref{GHZstates}.
In Sec.~\ref{Bell} the quantum protocol for a Bell-measurement is described in detail 
and numerically verified. We discuss  possible experimental implementations in  Sec.~\ref{discusion} and
our  conclusions are given  in Sec. \ref{conclusions}.

\section{The Two-photon model}
\label{model}
\begin{figure}[t]
  \includegraphics[width=.4\textwidth]{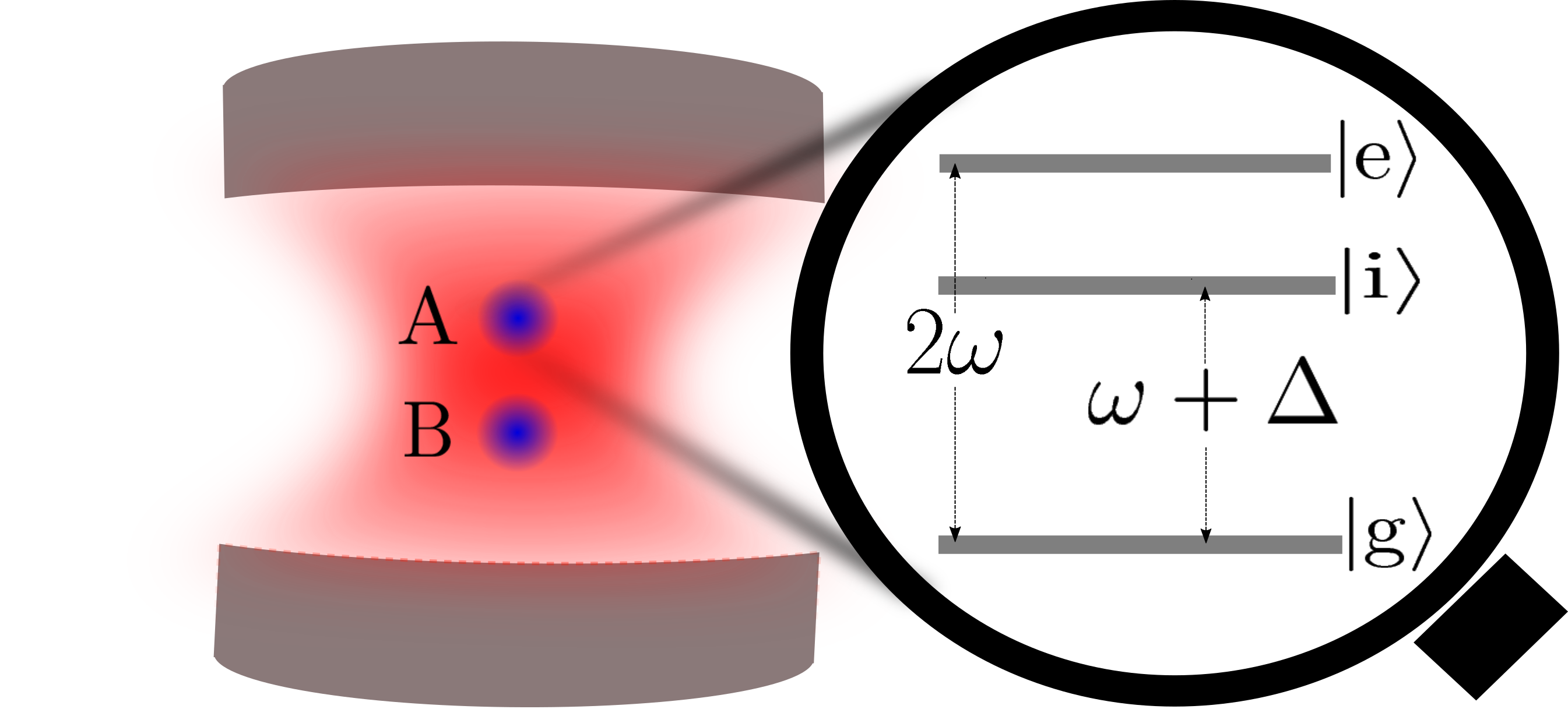}
  \caption{\label{levels} 
  Pictorial illustration of two three-level atoms (A and B) interacting at two-photon resonance 
  with one mode of the radiation field inside an optical cavity. For large enough detuning $\Delta$ 
  between the lower transition frequency and the frequency of the field, the intermediate 
  state is decoupled from the dynamics leading
  to an effective two-photon interaction involving only states $\ket{\rm g}$ and $\ket{\rm e}$ 
  of the atoms.
  }
\end{figure}

In this section we briefly review the Dicke model in the rotating wave approximation 
at two-photon resonance with two identical three-level atoms ($\rm{A}$ and $\rm{B}$) 
that interact with one mode of the quantized electromagnetic field inside an optical cavity.  
The field couples an intermediate level $\ket{\rm i}$ with the ground state $\ket{\rm g}$ and 
the excited state $\ket{\rm e}$ as depicted in Fig.~\ref{levels}. The frequency difference between 
ground and excited state is assumed to be tuned at twice the frequency of the cavity mode. 
Choosing units in which $\hbar=1$, the Hamiltonian describing the dynamics of the system can
be written as
\begin{equation}
  H=\omega a^\dagger a+2\omega S_{\rm ee}+(\omega+\Delta) S_{\rm ii}+V.
  \label{hamiltonian}
\end{equation}
The first term in the Hamiltonian describes the energy of the optical field and is written in terms
of the bosonic annihilation and creation operators $a$ and $a^\dagger$. The second and third term represent the atomic energy of the excited and intermediate states, respectively. They are expressed trough the atomic collective operators
\begin{equation}
  S_{\mu\nu}=
  \ketbra{\mu}{\nu}_{\rm A}
  +\ketbra{\mu}{\nu}_{\rm B}, \quad \mu,\nu\in\{ {\rm g}, {\rm i}, {\rm e}  \}.
  \label{eses}
\end{equation}
The last term in Eq.~\eqref{hamiltonian}, $V$, describes the atom-field interaction which is assumed to
fulfill the rotating-wave approximation (RWA) and therefore can be written as
\begin{equation}
  V=g_{\rm g}aS_{\rm ig} +g_{\rm e}aS_{\rm ei}+{\rm H.c.}
  \label{interaction}
\end{equation}
where $g_{\rm e}$ and $g_{\rm e}$ are the corresponding atom-field coupling strengths.

The detuning $\Delta$ between the frequency of the intermediate state and the frequency of the mode 
is assumed to be large compared with both coupling strengths that we consider  
of the same order of magnitude, namely $\Delta\gg g_{\rm g}\sim g_{\rm e}$.
In this particular situation, it can be shown that
the intermediate level can be approximately decoupled from the dynamics. 
To show this, we follow the method introduced in \cite{Klimov} and perform a small rotation of the Hamiltonian with the transformation  
\begin{equation}
e^{iG}He^{-iG},\quad
  G=
  \frac{g_{\rm g}}{\Delta}a S_{\rm ig}
  -\frac{g_{\rm e}}{\Delta}a S_{\rm ei}-{\rm H.c.}
  \label{}
\end{equation}
Using the Baker-Campbell-Hausdorff (BCH) expansion and neglecting terms of the order $g_{\rm e}(g_{\rm e}\sqrt{\langle a^\dagger a\rangle}/\Delta)^2$, one can obtain the following effective 
Hamiltonian
\begin{equation}
  H\approx\omega I+S+W,
  \label{hameff}
\end{equation}
which includes a two-photon interaction term
\begin{equation}
\label{interaction2}
W=g\left(a^2S_{\rm eg}+a^{\dagger\,2}S_{\rm ge}\right),
\quad 
g=-g_{\rm e}g_{\rm e}/\Delta.
\end{equation}
The expansion also produces a Stark-shift contribution of the form
\begin{equation}
  S=-2\frac{g_{\rm g}^2}{\Delta}I
  -\frac{g_{\rm e}^2-g_{\rm g}^2}{\Delta}aa^\dagger S_{\rm ee}
  +3\frac{g_{\rm g}^2}{\Delta}S_{\rm ee}.
  \label{stark}
\end{equation}
One can verify that the first term in Eq. \eqref{stark} is a constant of motion  
that is given by
\begin{equation}
  I=a^\dagger a+2S_{\rm ee}.
  \label{constant}
\end{equation}
In principle, $I$ should also contain the term $S_{\rm ii}$, 
however one can safely omit it if the intermediate state is not initially populated. 
The effective Hamiltonian in \eqref{hameff} can be verified with the commutation relations
$[G,S_{\rm ii}]=-V$ and  $[G,V]=2W+2S$ that follow from 
$[S_{\mu\nu},S_{\mu'\nu'}]=\delta_{\nu\mu'}S_{\mu\nu'}-\delta_{\nu'\mu}S_{\mu'\nu}$ using Eq. \eqref{eses}. Taking into account the order of the neglected terms, one can accurately describe
the dynamics of the system using the effective Hamiltonian subjected to the following 
restriction in time
\begin{equation}
g_{\rm e}t\ll\Delta^2/ g_{\rm e}^2\langle a^\dagger a\rangle.
\label{timecondition}
\end{equation}

In order to further simplify the interaction, one can find conditions for which the 
photon-dependent Stark-shift term, second in Eq.~\eqref{stark}, does not contribute to the dynamics. This part can be neglected if it is smaller 
than the omitted expressions in the truncated  BCH expansion leading to the effective Hamiltonian in Eq.~\eqref{hameff}, which  reduces to the condition 
$|g_{\rm e}^2-g_{\rm g}^2|< g_{\rm e}^3/\Delta $ quantifying the closeness between 
$g_{\rm g}$ and $g_{\rm e}$. With this in mind, the photon-independent Stark-shift, third in \eqref{stark}, is of the order of 
$g$, which can be neglected for large photon numbers compared with the order of $W$ given by 
$g\langle a^\dagger a\rangle$. Under these assumptions, one can reduce the Hamiltonian in Eq. \eqref{hamiltonian} simply to 
\begin{equation}
  H\approx 
  (\omega +2g)I+W.
  \label{Hamfin}
\end{equation}
As this Hamiltonian 
effectively describes the dynamics of the two atoms restricted to levels $\ket{\rm g}$
and $\ket{\rm e}$,
in what follows we will solve the Sch\"odinger equation for this Hamiltonian in the interaction
picture with respect to the constant of motion $(\omega+2g) I$ exploiting the fact that 
it commutes with the two-photon interaction, i.e., $[I,W]=0$.
We stress that under the aforementioned assumptions the dynamics of the system is well described by
the  two-photon interaction term $W$ in Eq.~\eqref{interaction2}.

\section{Solution of the dynamical equation for large photon number}
\label{solution}
In this section we derive an approximate analytical solution for the time-dependent state vector in the limit of large mean photon numbers.  To this end, we consider initial states of the form
$\ket\Psi=\ket\psi\ket\alpha$, where $\ket\psi$ is an arbitrary state of two two-level atoms, 
and where we have considered the photonic coherent state
\begin{equation}
  \ket\alpha=\sum_{n=0}^\infty p_n \ket n,
  \quad p_n=e^{-\frac{|\alpha|^2}{2}}\frac{\alpha^n}{\sqrt{n!}},
  \quad \alpha=|\alpha|e^{i\phi}.
  \label{alpha}
\end{equation}
The mean photon number is given by $\bar n=\langle a^\dagger a\rangle=|\alpha|^2$ 
	and in the following  it is  assumed to be large ($\bar n\gg 1$).
In order to find the time-dependent state vector we choose to solve the eigenvalue problem for $W$
using the photon number states $\ket n$, the atomic basis $\ket{\rm gg}$, $\ket{\rm ee}$,
and the Bell states
\begin{equation}
  \ket{\Psi^\pm}=\frac{1}{\sqrt2}\left(\ket{\rm ge}\pm\ket{\rm eg}\right).
  \label{}
\end{equation}
In this basis an arbitrary initial  state of the atoms takes the form
\begin{equation}
  \ket\psi=
  c_{\rm g}\ket{\rm gg}+
  c_-\ket{\Psi^-}+c_+\ket{\Psi^+}+
  c_{\rm e}\ket{\rm ee},
  \label{psi}
\end{equation}
where
the probability amplitudes fulfill the normalization condition and with
 the convention $\ket{\rm ge}=\ket{\rm g}_A\ket{\rm e}_B$.
It can be 
verified by inspection that the set of states $\{\ket{\Psi^-}\ket{n}\}_{n=0}^\infty$ 
are eigenvectors of $W$ with eigenvalue $0$. The rest of the eigensystem can be evaluated
by diagonalizing $3\times 3$ matrices which
in the tripartite  basis $\{\ket{\rm gg}\ket{n},\ket{\Psi^+}\ket{n-2},\ket{\rm ee}\ket{n-4}\}$, 
can be written as
\begin{eqnarray}
  W_n&=&g\sqrt2\nonumber\left(
  \begin{array}{ccc}
    0 & \sqrt{n^2-n} & 0\\
    \sqrt{n^2-n}&0&\sqrt{n^2-5n+6}\\
  0&\sqrt{n^2-5n+6}&0
  \end{array}
  \right).
  \nonumber
  \label{interaction3}
\end{eqnarray}
Although it is possible to diagonalize these matrices in an exact form, 
the condition of high mean photon number $|\alpha|^2\gg 1$ will allow us
to find compact expressions that are good approximations to the exact results. 
For instance, the exact nonzero eigenvalues are 
$w_{n}^\pm=\pm g\sqrt{(2n-3)^2+3}$, but they can be approximated for large values of
$n$ by
\begin{equation}
  \tilde w_n^\pm=\pm g(2n-3).
  \label{}
\end{equation}
In this limit, one can find that the orthogonal transformation which diagonalizes each block
$W_n$ takes the simple form
\begin{equation}
  \tilde O_n=\frac{1}{2}\left(
  \begin{array}{ccc}
    -\sqrt2 & 1 & 1\\
    0&-\sqrt2&\sqrt2\\
    \sqrt2&1&1
  \end{array}
  \right).
  \label{}
\end{equation}
The evolution operator can also be expressed in terms of matrices of size $3\times 3$,
which can be evaluated using the transformation that diagonalizes the blocks $W_n$ of
$W$, namely 
\begin{equation} 
  \tilde U_n(t)=\tilde O_n^\intercal\exp[-i\, {\rm diag}(0,-\tilde w_n,\tilde w_n) t]\tilde O_n,
  \label{}
\end{equation}
where ${{\rm diag} (v)}$ represents a diagonal matrix with the elements of $v$ as non-zero entries. 
With these approximations, the evolution operator has  the remarkable simple form
\begin{equation}
  \tilde U_n(t)=\left(
  \begin{array}{ccc}
    \cos^2(\frac{\tilde w_nt}{2}) & \frac{\sin(\tilde w_nt)}{i\sqrt2} & -\sin^2(\frac{\tilde w_nt}{2}) \\
    \frac{\sin(\tilde w_nt)}{i\sqrt2}&\cos(\tilde w_n t)&\frac{\sin(\tilde w_nt)}{i\sqrt2}\\
    -\sin^2(\frac{\tilde w_nt}{2})&\frac{\sin(\tilde w_nt)}{i\sqrt2}&\cos^2(\frac{\tilde w_nt}{2}) 
  \end{array}
  \right).
  \label{}
\end{equation}
Using these results, one can find that the time evolution of any initial state can be written as
\begin{align}
  \label{Psit}
  \ket{\Psi(t)}&=a_0\ket{{\rm gg},0}+a_{1}\ket{{\rm gg},1}+
  \nonumber
  \\&
  \sum_{n=2}^3\left( a_{n,t}\ket{{\rm gg},n}+b_{n,t}\ket{\Psi^+,n-2}\right)+
  \\&
  \sum_{n=4}^\infty \left(a_{n,t}\ket{{\rm gg},n}+b_{n,t}\ket{\Psi^+,n-2}+c_{n,t}\ket{{\rm ee},n-4}\right),
  \nonumber
\end{align}
where $a_0=c_{\rm e}p_0$ and $a_1=c_{\rm e}p_1$ are the  probability amplitudes of stationary states that are decoupled from the dynamics. 
The rest of the coefficients can be evaluated with the aid of the evolution operator and are given by
\begin{align}
  a_{n,t}&=
  \left(
  \tfrac{c_++d_{2\phi}^+}{2}e^{-i\tilde w_n t}
  -\tfrac{c_+-d_{2\phi}^+}{2}e^{i\tilde w_n t}+d_{2\phi}^-
  \right)\tfrac{e^{-i2\phi}p_n}{\sqrt2},
  \nonumber
  \\
  b_{n,t}&=
  \left(
  \tfrac{c_++d_{2\phi}^+}{2}e^{-i\tilde w_n t}
  +\tfrac{c_+-d_{2\phi}^+}{2}e^{i\tilde w_n t}
  \right)p_{n-2},
  \\
  c_{n,t}&=
  \left(
  \tfrac{c_++d_{2\phi}^+}{2}e^{-i\tilde w_n t}
  -\tfrac{c_+-d_{2\phi}^+}{2}e^{i\tilde w_n t}-d_{2\phi}^-
  \right)\tfrac{ e^{i2\phi}p_{n-4}}{\sqrt2}.
  \nonumber
  \label{coefficients}
\end{align}
In the previous expressions we have introduced for notational convenience the coefficients 
\begin{equation}
  d_\phi^\pm=\frac{c_{\rm g} e^{i\phi}\pm c_{\rm e}e^{-i\phi}}{\sqrt2},
\end{equation}
which are the initial probability amplitudes of the maximally entangled
states of the two atoms 
\begin{equation}
  \ket{\Phi_\phi^\pm}=\frac{1}{\sqrt2}\left(e^{-i\phi}\ket{\rm gg}\pm e^{i\phi}\ket{\rm ee}\right).
  \label{}
\end{equation}

\begin{figure}[t]
\includegraphics[width=.43\textwidth]{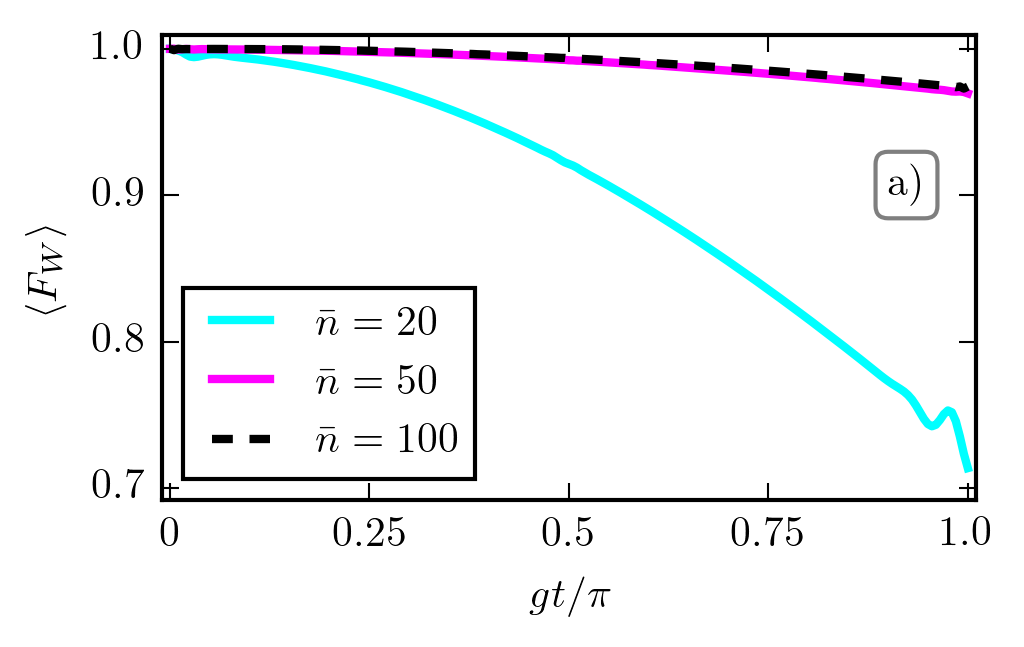}
\includegraphics[width=.43\textwidth]{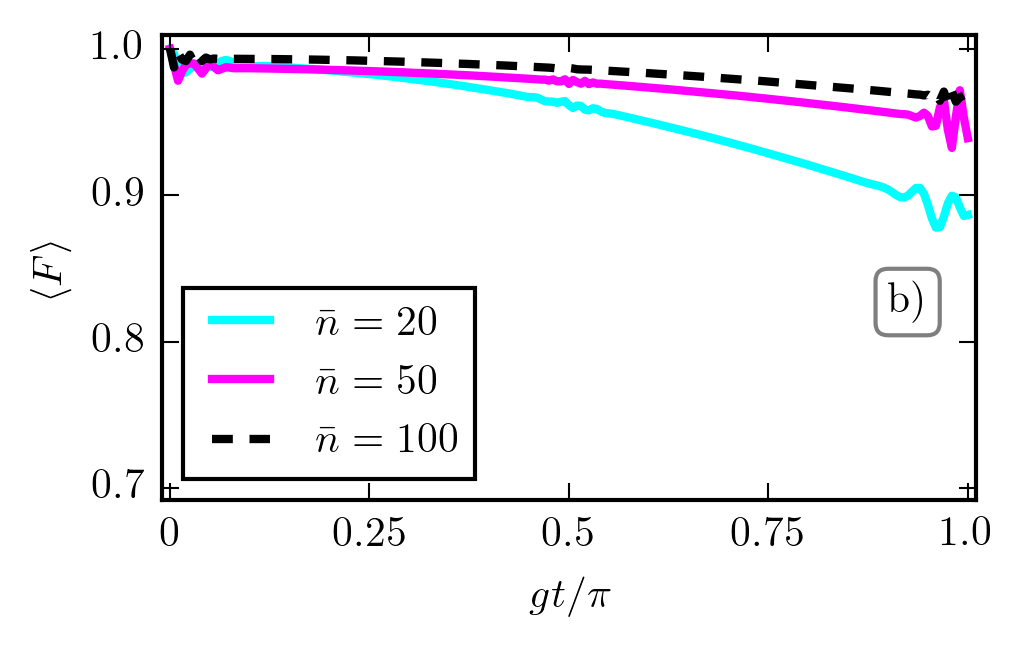}
\caption{\label{fidfig}
a) Ensemble average of the fidelity $\langle F_{W}\rangle$ computed from the exact numerical state vector $\ket{\Psi_{W}(t)}$ for the effective two-photon Hamiltonian $W$ respect to the numerical solution $\ket{\Psi_{\rm exact}(t)}$ corresponding to the full Hamiltonian \eqref{hamiltonian} as a function of the scaled time $gt/\pi$. Three cases are shown: $\bar{n}=20, 50, 100$.  
b) Ensemble average of the fidelity $\langle F\rangle$ of the approximate solution $\ket{\Psi(t)}$ (Eq. \eqref{Psiap}) also with reference to the state $\ket{\Psi_{\rm exact}(t)}$ as function of the scaled time $gt$ with  $g_{\rm{g}}/\Delta=0.002$. 
The ensemble average has been performed over $10^3$ initial random pure states uniformly distributed according to the Haar measure of $SU(4)$ with $\phi$ a random phase 
drawn from a uniform distribution $\in$ $[0,2\pi)$.
}
\end{figure}

In order to find a simple expression for the state vector, we use the following approximate
relation for the photonic probability amplitudes
\begin{equation}
  p_n\approx p_{n-1}e^{i\phi}.
  \label{pnapp}
\end{equation}
With this result one can carry out the summation in Eq. \eqref{Psit} in order the  arrive to an approximation of the state vector in terms of coherent states and maximally
entangled atomic states, namely
\begin{align}
 \label{Psiap}
  \ket{\Psi(t)}&=
  \left(c_-\ket{\Psi^-}+d_{2\phi}^-\ket{\Phi_{2\phi}^-}\right)\ket\alpha+
  \\
  &\frac{c_++d_{2\phi}^+}{2} e^{-igt}\left(\ket{\Psi^+}+\ket{\Phi_{2\phi-4gt}^+}\right)
  \ket{e^{-i2 gt}\alpha}+
  \nonumber
  \\
  &\frac{c_+-d_{2\phi}^+}{2}e^{igt}\left(\ket{\Psi^+}-\ket{\Phi_{2\phi+4gt}^+}\right)
  \ket{e^{i 2gt}\alpha}.
  \nonumber
\end{align}
A similar expression to this formula was found in \cite{Torres2016} for the two-atom 
Tavis-Cummings model involving more complicated field states that followed the dynamics
of a coherent state, but distorting its shape in time. In contrast, the solution in \eqref{Psiap} is 
written in terms of coherent states as a consequence of the linear behavior of the eigenfrequencies 
of the system for large photon numbers. The full exact solution for the two-photon model
was previously reported in Ref. \cite{Bashkirov} together with a semiclassical approximation in 
agreement with our findings. There, however,  
the form in terms of orthogonal Bell states and coherent states was not identified nor its
potential application was stressed.

In order to test the validity of our approximation, we have plotted in Fig.~\ref{fidfig}~a) the  fidelity
$F_{W}=|\braket{\Psi_{\rm {exact}}(t)}{\Psi_{W}(t)}|^2$ of the exact numerical solution evaluated with 
the full Hamiltonian in Eq.~\eqref{hamiltonian}
with respect to the exact numerical state vector computed with the two-photon Hamiltonian
in \eqref{Hamfin} for different values of the average photon number $\bar n$. 
As a comparison we show  in Fig.~\ref{fidfig}~b) the fidelity $F=|\braket{\Psi_{\rm exact}(t)}{\Psi(t)}|^2$ with respect to the approximate state vector in terms of coherent states 
of Eq.~\eqref{Psiap}. 
In favor of generality, and for both fidelities, we have performed an ensemble average with $10^3$ random initial pure states taken from the uniform distribution of $SU(4)$. The phase $\phi$ of the coherent state
was randomly obtained from a uniform distribution in the interval $[0,2\pi)$. It can be noted, 
that the agreement between dynamics is remarkably good for increasing value of the mean photon number in both situations.  Having checked its validity, the solution in Eq.~\eqref{Psiap} will be the starting point of our subsequent analysis. 

\section{Collapse and revival of Rabi oscillations}
\label{Rabi}

\begin{figure}[t]
    \includegraphics[width=.45\textwidth]{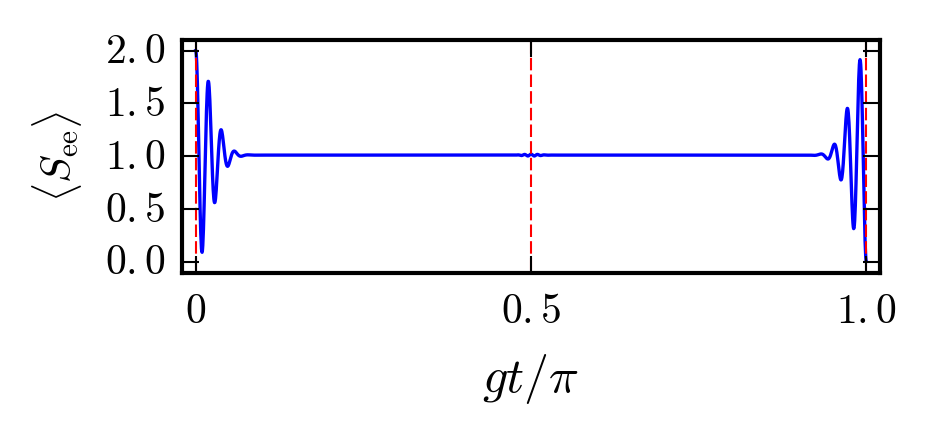}
  \caption{\label{revivialfig}
 Collapse and perfect revival of Rabi oscillations in $\langle S_{\rm ee}\rangle$ 
  		(number of atoms in the excited state)
  as a function of the scaled time $gt/\pi$.
 The initial state $\ket{\text{ee}}\ket{\alpha}$ with $|\alpha|^2=50$ evolves
 under the influence of the effective Hamiltonian \eqref{Hamfin}. 
 An approximate analytical expression for these oscillations is given in Eq. \eqref{RabiSee}.
  }  
\end{figure}

A clear manifestation of the coherent shape of the components of the field state is the perfect revivals
of the Rabi oscillations of observables such as the mean value of the operator $S_{\rm ee}$, which can be interpreted as the number of atoms in their corresponding excited state. This can be evaluated analytically, for instance for the initial state
$\ket{\text{ee}}\ket{\alpha}$, using our expression \eqref{Psiap} as
\begin{equation}
\label{RabiSee}
\left\langle S_{\rm ee}\right\rangle= 1+\operatorname{Re}
\left[e^{-|\alpha|^2(1-e^{i2gt})-i3gt}\right],
\end{equation}
where we have  employed 
the overlap between the relevant coherent states
$\braket{\alpha}{\alpha e^{\pm i2gt}}=e^{-|\alpha|^2(1-e^{\pm i2gt})}$
which has a Gaussian envelope  
$(1+e^{-2|\alpha|^2 g^2 t^2})/2$ 
for values of time close to zero and in general to $gt=\pi l$, with $l\in\mathbb{N}$.
In Fig.~\ref{revivialfig} we have plotted the numerical exact calculation of 
$\left\langle S_{\rm ee}\right\rangle$.
As in the case of the standard Jaynes-Cummings interaction, collapses and revivals in this atomic observable are  present in the dynamics of the two-photon model (apart from an alternating sign). However, they show a different behavior as they appear in a more compact and regular form,
showing almost the complete returning to the initial photonic state in the case of large fields \cite{Puri88}. In the two-photon two-atom model,
the time at which revivals appear is independent of  $\bar n$ and is  given by 
\begin{equation}
t_{r}\approx \pi/g.
\label{trevival}
\end{equation}
In order to attain $t_r$ with the model of Sec. \ref{model}, the restriction in time of 
Eq. \eqref{timecondition} results in the following condition for
the parameters of  the model:  $g_{\rm e}\bar n\pi\ll \Delta$.

\begin{figure}[t]
    \includegraphics[width=.48\textwidth]{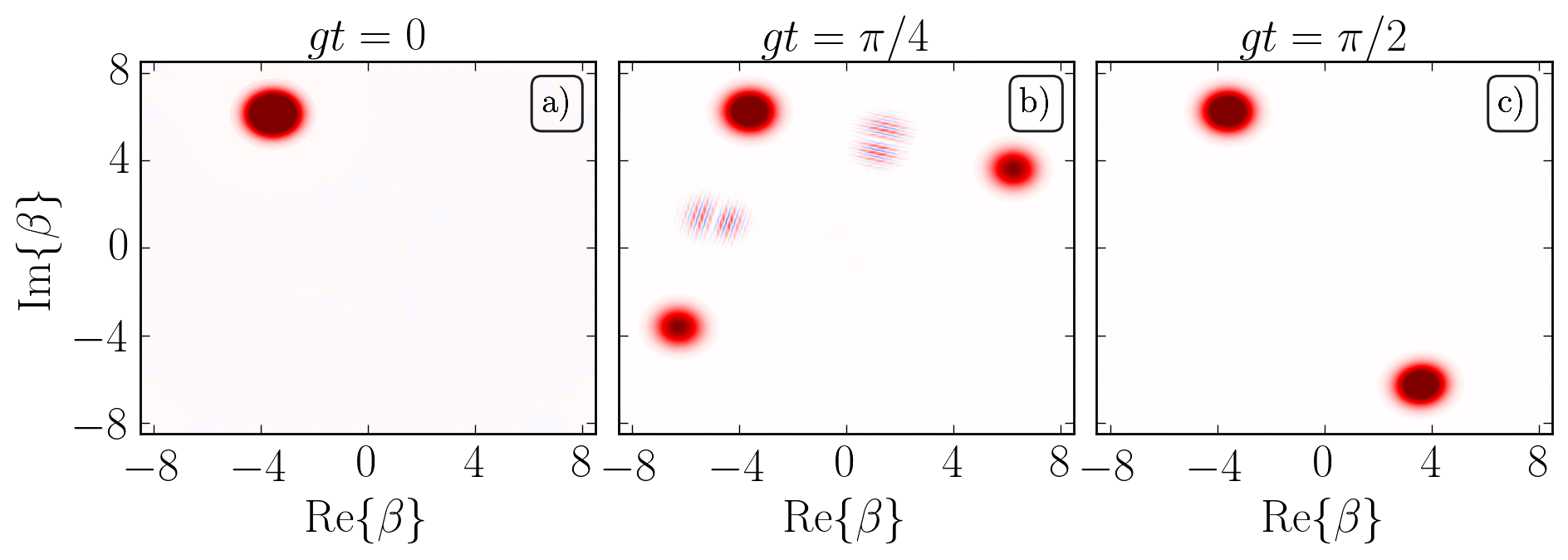}
  \caption{\label{wigfig}
Wigner function of the optical field for interactions times a) $t_{r}=0$, b) $t_{r}=1/4$, c) $t_{r}=1/2$.
  At $t_{r}=1/2$ (last snapshot)  one can recognize the shape of an incoherent superposition of two coherent states. 
  Parameters are the same as in Fig. \ref{revivialfig} with $\phi=2\pi/3$.
  }
\end{figure}

The collapse and revival of Rabi oscillations can also be studied in phase space. 
This gives a relevant pictorial description of the time-evolution of the field state, whose form
will corroborate our approximation in terms of coherent states. We choose to visualize the behavior
in terms of the Wigner function \cite{Vogel}, a quasi-probability distribution defined as 
\begin{equation}
 W(\beta,t)=\frac{1}{\pi^2}\int\text{Tr}\left\{\rho_{f}(t)e^{\zeta a^{\dagger}-\zeta^{*}a}\right\}e^{\beta\zeta^{*}-\beta^{*}\zeta}d^{2}\zeta,
 \label{wigner}
\end{equation}  
with $\beta$ and $\zeta$ being complex numbers and the reduced density operator 
of the field obtained after tracing out the atomic degrees of freedom, i.e., $\rho_{f}(t)={\rm{Tr}}_{A,B}\ket{\Psi(t)}\bra{\Psi(t)}$. 
In Fig.~\ref{wigfig} we present the Wigner function of the photonic state for three different values of the
interaction time, namely $t=0$, $t=t_r/4$ and $t=t_r/2$. 
From this representation one can extract relevant dynamical information of the full system. 
The initial state for $t=0$ in Fig.~\ref{wigfig}~a) corresponds to a coherent state and is represented by a Gaussian distribution in
the complex plane. For nonzero values of the interaction time, the field evolves as correlated
coherent states, without deforming its circular shape, showing  no squeezing during the evolution. The correlated feature is manifested by the interference fringes between the maxima at $t_r/4$ in 
Fig.~\ref{wigfig}~b) that disappear at $t_r/2$ in Fig.~\ref{wigfig}~c). From Eq. \eqref{Psiap}, one can evaluate the
state vector at  half the revival time
\begin{align}
  \ket{\Psi(t_r/2)}&=
  \left(c_-\ket{\Psi^-}+d_{2\phi}^-\ket{\Phi_{2\phi}^-}\right)\ket\alpha
  \nonumber
  \\
  &-i\left(c_+\ket{\Phi_{2\phi}^+}+d_{2\phi}^+\ket{\Psi^+}\right)\ket{-\alpha}.
  \label{Psiatt2}
\end{align}
Tracing over the atomic degrees of freedom, one finds 
that the field state corresponds to the mixed state 
\begin{align}
\rho_{f}=&\left(|c_{-}|^{2}+|d^{-}_{2\phi}|^{2}\right)\proj{\alpha}+\nonumber\\
&\left(|c_{+}|^{2}+|d^{+}_{2\phi}|^{2}\right)\proj{-\alpha}.
\end{align}
This incoherent superposition explains the absence of interference fringes between the two-dimensional Gaussian functions representing opposed coherent states in phase space in  Fig.~\ref{wigfig}~c). The complete state of the system at $t_r/2$, half revival time, 
given in Eq. \eqref{Psiatt2} will play a key role in what follows. 
In the next section we will show that multipartite quantum correlations can be generated during the time evolution leading to the formation of tripartite entangled states. 

\section{Generation of GHZ states}
\label{GHZstates}
An immediate application is
the possibility to generate maximally entangled three-qubit states
using the intrinsic dynamics of the two-photon model. 
Based on our solution in terms of Bell and coherent states at half revival time
in Eq.~\eqref{Psiatt2} and setting the coefficients $c_{-}=d^{+}_{2\phi}=0$, and $c_{+}=\frac{1}{\sqrt{2}}$, $d^{-}_{2\phi}=\frac{i}{\sqrt{2}}$,
the state vector evaluated at $t_r/2$ takes the following form:
\begin{align}
  \ket{\Psi(t_r/2)}=\frac{i}{\sqrt{2}}
  \left[\ket{\Phi_{2\phi}^-}\ket\alpha-\ket{\Phi_{2\phi}^+}\ket{-\alpha}\right].
  \label{statechido}
\end{align}
Looking at the probability amplitudes, the initial state might appear somehow complicated or even entangled, but it is actually an initial tripartite product state with the field in the coherent state
$\ket\alpha$ and each atom in the state
\begin{align}
  \ket{\varphi_\phi}=\frac{e^{i\pi/4}}{\sqrt2}\left(e^{-i\phi}\ket{\rm g}-ie^{i\phi}\ket{\rm e}\right).
  \label{statechidocero}
\end{align}
It is therefore a remarkable result that the simple unitary evolution generates a maximally entangled
tripartite state with a product state as an input. In order to show that this corresponds to a tripartite
entangled state, it is useful to establish an isomorphism between coherent and qubit states for large values
of $|\alpha|$.  Consider the following even and odd Schr\"odinger cat states:
\begin{align}
\ket{\alpha,\pm}=\frac{1}{\sqrt{2}}\left(\ket{\alpha}\pm\ket{-\alpha}\right),
  \label{catstates}
\end{align}
which are eigenstates of the parity operator $\Pi=(-1)^{a^{\dagger}a}$ with eigenvalues $\pm 1$ that
fulfill the condition $\braket{\alpha,+}{\alpha,-}\approx 0$ for $|\alpha|\gg 1$.
Even and odd cat states can then be respectively interpreted as the excited and ground states of a two-level system \cite{Gerry96}.  In fact, one can easily check that the operators: 
\begin{align}
\Sigma _{x}&=\ket{\alpha,+}\bra{\alpha,-}+\ket{\alpha,-}\bra{\alpha,+}, \\
\Sigma _{y}&=i\left(\ket{\alpha,+}\bra{\alpha,-}-\ket{\alpha,-}\bra{\alpha,+}\right),\\
\Sigma _{z}&=\ket{\alpha,+}\bra{\alpha,+}-\ket{\alpha,-}\bra{\alpha,-},
\end{align}
satisfy the same $SU(2)$ algebra $[\Sigma_{i},\Sigma_{j}]=2i\epsilon_{ijk}\Sigma_{k}$ and 
$\{\Sigma_{i},\Sigma_{j}\}=\delta_{ij}$. If we set the phase $\phi=\pi/4$, the state in Eq.~\eqref{statechido} can be rewritten as 
\begin{align}
\ket{\rm{GHZ}}=\frac{1}{\sqrt{2}}\left(\ket{\rm {gg}}\ket{\alpha,-}+\ket{\rm{ee}}\ket{\alpha,+}\right),
  \label{GHZstate}
\end{align}
which can be immediately  recognized as a Greenberger-Horne-Zeilinger (GHZ) state. 
It is well known that GHZ states contain one of the two types of tripartite entanglement, and they have been 
experimentally realized in a variety of physical systems, such as photons, trapped ions, and superconducting qubits \cite{Zeilinger99,Monz11,kelly2015}. 
These states are characterized by the fact that a measurement performed on the third qubit results in an unentangled qubit pair. 
\begin{figure}[t]
  \includegraphics[scale=1.]{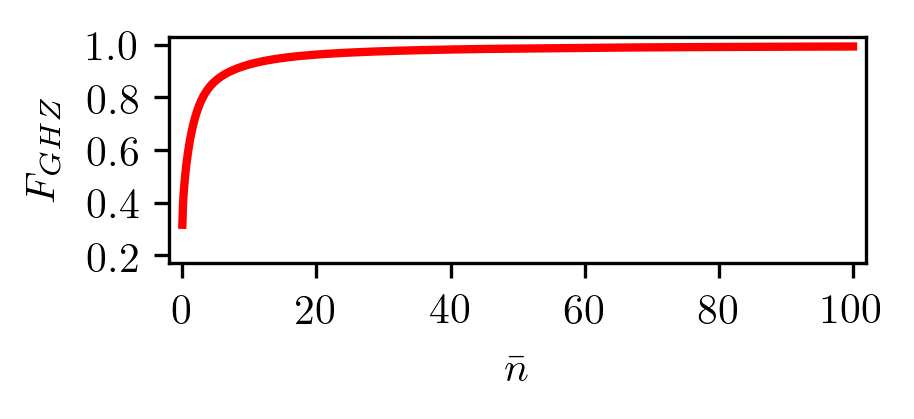}
  \caption{\label{ghzfidelity}
Fidelity between the generated GHZ state in Eq.~\eqref{GHZstate} 
and the numerically evaluated state vector, $F_{GHZ}=|\braket{\rm GHZ}{\Psi_{\rm exact}(\pi/2g)}|^2$, as a function of the mean photon number.}
\end{figure}
However, 
a very interesting fact is that pairwise entanglement can be obtained by performing an appropriate measurement of the third qubit along some
orthogonal direction. From Eq.~\eqref{GHZstate} we can see this by projecting onto the coherent states $\ket{\pm\alpha}$, which automatically leaves the qubit pair in one of the entangled states $\ket{\Phi^{\pm}}$. 
The corresponding fidelity $F_{GHZ}=|\braket{\rm GHZ}{\Psi_{\rm exact}(\pi/2g)}|^2$ between the generated GHZ state in Eq.~\eqref{GHZstate} and the exact state vector calculated by numerical means is shown in Fig.~\ref{ghzfidelity}
for increasing  mean photon number. As we have shown, almost unit fidelity GHZ states can be efficiently engineered by the appropriate tuning of the initial conditions.

\section{Bell measurement} 
\label{Bell}
\begin{figure}[t]
  \includegraphics[width=.48\textwidth]{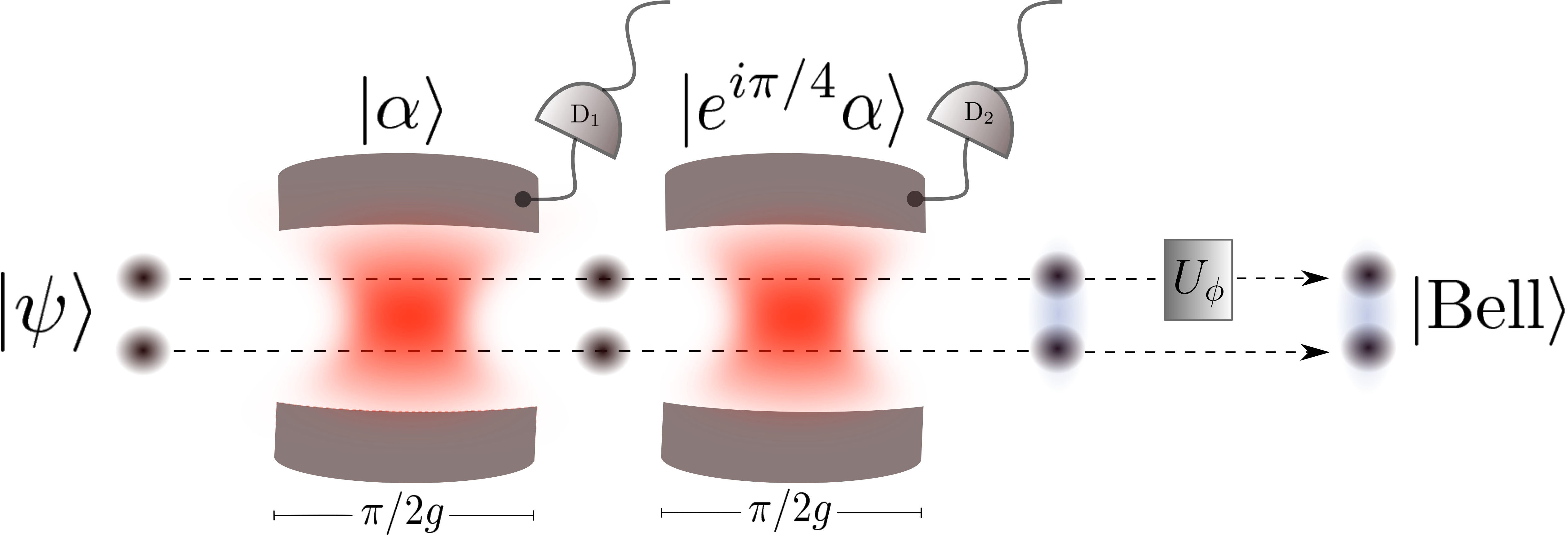}
  \caption{\label{protocolfig}
 Schematic visualization of the Bell measurement scheme. Two matter
 qubits initially described by an arbitrary state $\ket{\psi}$ enter to a sequence of two independent electromagnetic cavities prepared in coherent states. Both qubits couple to each mode for a time $t_{r}/2$. At the exit of the cavities, a measurement on the field state is performed by detectors D$_1$ and D$_2$. As a final step, a single-qubit unitary $U_{\phi}$ is applied on the first qubit resulting in the postselection of a Bell state.}
\end{figure}

In this section we present a scheme to implement a complete Bell measurement based on 
atomic postselection by letting the atoms interact with two separate cavities and
then measuring the field state as depicted in Fig.~\ref{protocolfig}.
We elucidate this by first considering the interaction with one cavity for an interaction 
time equal to half revival time for which the system is left in the state given by Eq. \eqref{Psiatt2}.
At this time, only two coherent states contribute to the photonic state and are correlated with
two orthogonal components of the atomic state. As we are considering the limit of high excitation
number, these two coherent states are nearly orthogonal as can be noted from their overlap
$\braket{\alpha}{-\alpha}=e^{-2|\alpha|^2}$, and therefore  can be distinguished with an
appropriate measurement scheme \cite{Leuchs,Torres2014,Torres2016}. 
For our discussion, we assume that one is able to project the field  onto the states  $\ket\alpha$ or $\ket{-\alpha}$.
From the form of the state in Eq. \eqref{Psiatt2}, one can note that  these projections
correspond, respectively, to the following atomic measurement operators
\begin{align}
   M^+_{\phi}&=\proj{\Psi^-}+\proj{\Phi^-_{2\phi}},
  \nonumber\\
   M^-_{\phi}&=-i\ketbra{\Phi^+_{2\phi}}{\Psi^+}-i\ketbra{\Psi^+}{\Phi^+_{2\phi}}.
  \label{projat}
\end{align}
Therefore, projecting onto the photonic state $\ket{\pm\alpha}$ corresponds to implementing  the atomic
measurement operator $M^\pm_{\phi}$  composed of a rank-two projector and a flip operator, both given in the
Bell basis.
The appearance of this flip operator  is due to the fact that in \eqref{Psiatt2}
the initial atomic probability amplitudes of the state in the second row are interchanged. 
For certain atomic probability amplitudes, the above projection postselect the atoms in an entangled state in a similar fashion as in \cite{Bernad2016}.

With the previous result it is not possible to project the atomic states into four orthogonal states,
as we have only encountered a rank-two projector and a flip operation in the space
of two maximally entangled states. However, one can  extend this result with 
the use of a second cavity, similar to \cite{Torres2014,Torres2016}. For this purpose, one has to let the
atoms interact with the field  prepared in a coherent state of the form $\ket{e^{i\pi/4}\alpha}$, i.e.,  
phase-shifted by $\pi/4$ from the first coherent state. This can be done, for instance, by letting
the atoms interact with a second cavity as depicted in Fig.~\ref{protocolfig}. After an interaction 
time of $t_r/2$, one would obtain a similar state to the one in Eq.~\eqref{Psiatt2}, but with rotated coherent states, i.e., $\phi$
replaced by $\phi+\pi/4$. In this case, projecting onto $\ket{\pm e^{i\pi/4}\alpha}$ would correspond
to measuring the atoms with a measurement operator $M^\pm_{\phi+\pi/4}$. Combining this with the previous procedure, one is able to measure the atoms according to the following measurement elements
\begin{align}
  \nonumber
  M^{++}= M^+_{\phi+\pi/4} M^+_{\phi}&=\proj{\Psi^-},
  \\
  \nonumber
  M^{+-}= M^+_{\phi+\pi/4} M^-_{\phi}&=-i\ketbra{\Phi^+_{2\phi}}{\Psi^+},
  \\
  M^{-+}= M^-_{\phi+\pi/4} M^+_{\phi}&=\ketbra{\Psi^+}{\Phi^-_{2\phi}},
  \\
  \nonumber
  M^{--}= M^-_{\phi+\pi/4} M^-_{\phi}&=i\ketbra{\Phi^-_{2\phi}}{\Phi^+_{2\phi}}.
  \label{emes}
\end{align}
where we have  considered the following  relations
\begin{equation}
  \ket{\Phi_{\phi+\pi/2}^\pm}=-i\ket{\Phi^\mp_{\phi}}, \quad d^\pm_{\phi+\pi/2}=id^\mp_{\phi}.
  \label{pihalf}
\end{equation}
Each measurement element $M^{\pm\pm}$ corresponds to the simultaneous projection onto 
$\ket{\pm\alpha}$ in the first cavity and onto $\ket{\pm e^{i\pi/4}\alpha}$ in the second cavity.
It turns out that all  $M^{\pm\pm}$ form the set of measurement operators of a particular
positive operator-valued measurement (POVM) 
\cite{Nielsen} that is already good enough to distinguish the four  Bell states. However, this does
not correspond to a von Neumann measurement, as there are some states that are interchanged during
the process. In order to convert this scheme into a von Neumann measurement of the four Bell
states, i.e., a Bell measurement, one has to implement a procedure to flip some of the Bell states.
Fortunately, this can be accomplished with the help of the following pair of single-atom unitary transformations
\begin{equation}
  \sigma_\phi=e^{i\phi}\ketbra{\rm e}{\rm g} +e^{-i\phi}\ketbra{\rm g}{\rm e},\quad
  \sigma_z=\ketbra{\rm e}{\rm e} -\ketbra{\rm g}{\rm g},
  \label{gates}
\end{equation}
that transform the Bell states according to the following rules
\begin{align}
\sigma_{\phi,A}\ket{\Phi_\phi^\pm}=\pm\ket{\Psi^\pm},\quad 
\sigma_{z,A}\ket{\Psi^\pm}=- \ket{\Psi^\mp}\nonumber\\
\sigma_{\phi,A}\ket{\Psi^\pm}=\pm\ket{\Phi^\pm_\phi},\quad 
\sigma_{z,A}\ket{\Phi^\pm_\phi}=- \ket{\Phi^\mp_\phi}.
\end{align}
Applying these single-qubit gates only to qubit-$A$ in a selective way after the field measurement 
and according to each outcome, one is able to perform the following Bell-state projections
\begin{align}
  \nonumber
  M^{++}=&\proj{\Psi^-},
  \\  
  \nonumber
  i\sigma_{2\phi,A}M^{+-}=&\ketbra{\Psi^+}{\Psi^+},
  \\
  \nonumber
  \sigma_{2\phi,A}\sigma_{z,A}M^{-+}=&\ketbra{\Phi^-_{2\phi}}{\Phi^-_{2\phi}},
  \\
  i\sigma_{z,A}M^{--}=&\ketbra{\Phi^+_{2\phi}}{\Phi^+_{2\phi}},
  \label{emes2}
\end{align}
that are required in a complete Bell measurement.
The implementation of the selective single-qubit gate is represented in Fig.~\ref{protocolfig} by the application of the operation $U_{\phi}$
after the interaction with the two cavities. A summary of the protocol with the corresponding single-qubit gate on atom $A$ is presented in table \ref{table}.
\begin{table}[t]
  \begin{tabular}{cccc}
    \hline\hline
    Measured field \,\,& Measured field\,\,  & Postselected\,\, & Gate $U_\phi$ \\
     state in $\rm{D_{1}}$ & state in $\rm{D_{2}}$ & Bell state&  on qubit $A$\\
    $\ket{\alpha}$ & $\ket{e^{\rm{i}\pi/4}\alpha}$ & $\ket{\Psi^{-}}$ & $\mathbbm{1}$ \\
     $\ket{\alpha}$ & $\ket{-e^{\rm{i}\pi/4}\alpha}$ & $\ket{\Psi^{+}}$ & $i\sigma_{2\phi}$ \\
     $\ket{-\alpha}$ & $\ket{e^{\rm{i}\pi/4}\alpha}$ & $\ket{\Phi^{-}_{2\phi}}$ & $\sigma_{2\phi}\sigma_z$ \\
     $\ket{-\alpha}$ & $\ket{-e^{\rm{i}\pi/4}\alpha}$ & $\ket{\Phi^{+}_{2\phi}}$ & $i\sigma_z$ \\
    \hline\hline
  \end{tabular}
  \caption{\label{table} Summary of the quantum protocol indicating the measured field in each cavity, the corresponding post-selected Bell state and the unitary gate that has to be applied to complete
  a Bell measurement.}
\end{table}

\begin{figure}[t]
  \includegraphics[width=.46\textwidth]{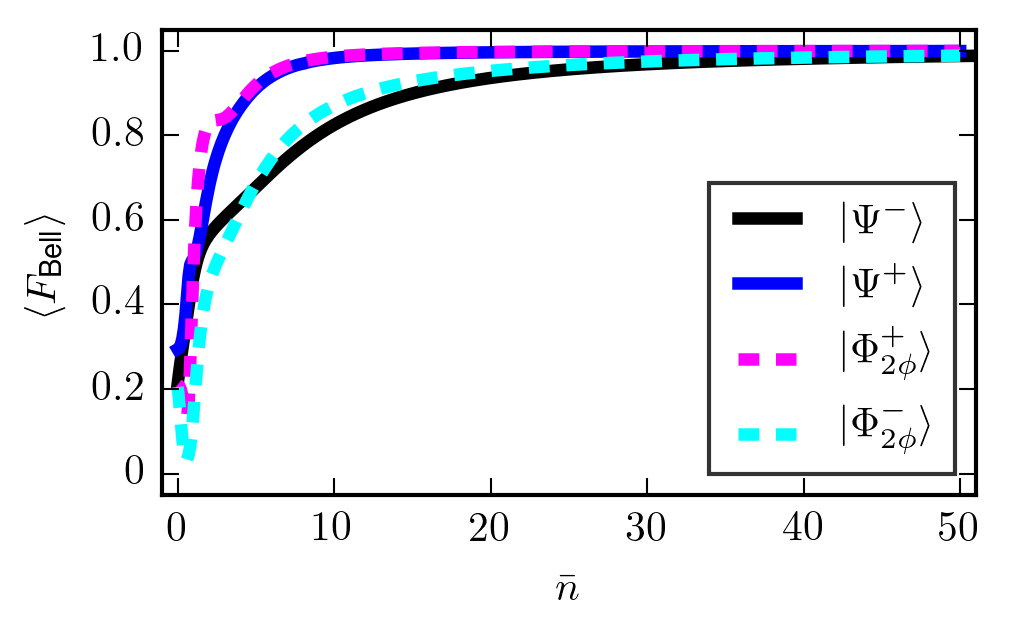}
  \caption{\label{bellfidfig}
Ensemble average fidelity $\langle F_{\rm Bell}\rangle$ for each postselected Bell state after the application of the protocol  as a function of the mean photon number.}
\end{figure}

In order to test the protocol, we have carried out numerical simulations to evaluate the fidelity of the
postselected atomic states with respect to the corresponding Bell state, i.e., 
$F_{\rm{Bell}}=\bra{\psi_{\rm {Bell}}}\rho_{\rm at}\ket{\psi_{\rm {Bell}}}$, 
where $\ket{\psi_{\rm{Bell}}}$ stands for one of the four Bell states and $\rho_{\rm at}$ is the reduced
density matrix of the two atoms after implementation of the protocol in Fig.~\ref{protocolfig}.
We have also performed an average over $10^3$ initial atomic random pure states from a uniform $SU(4)$ distribution in order to produce generic results for two-qubit systems. 
In Fig.~\ref{bellfidfig} we plot the  average fidelity of the numerically obtained
Bell states as a function of the mean photon number. We can see that even for relatively small photon numbers, a complete  Bell-measurement 
can be implemented following the proposed protocol. 
In this case, the only requirement for  
the mean photon number is to be sufficiently large. This contrasts with previous findings  based on the Tavis-Cummings model for which the fidelity of two of the four Bell states is an oscillatory function of the mean photon number, making the protocol functional only for restricted values of the mean photon number \cite{Torres2016}. 

\begin{figure}[t]
  \includegraphics[width=.46\textwidth]{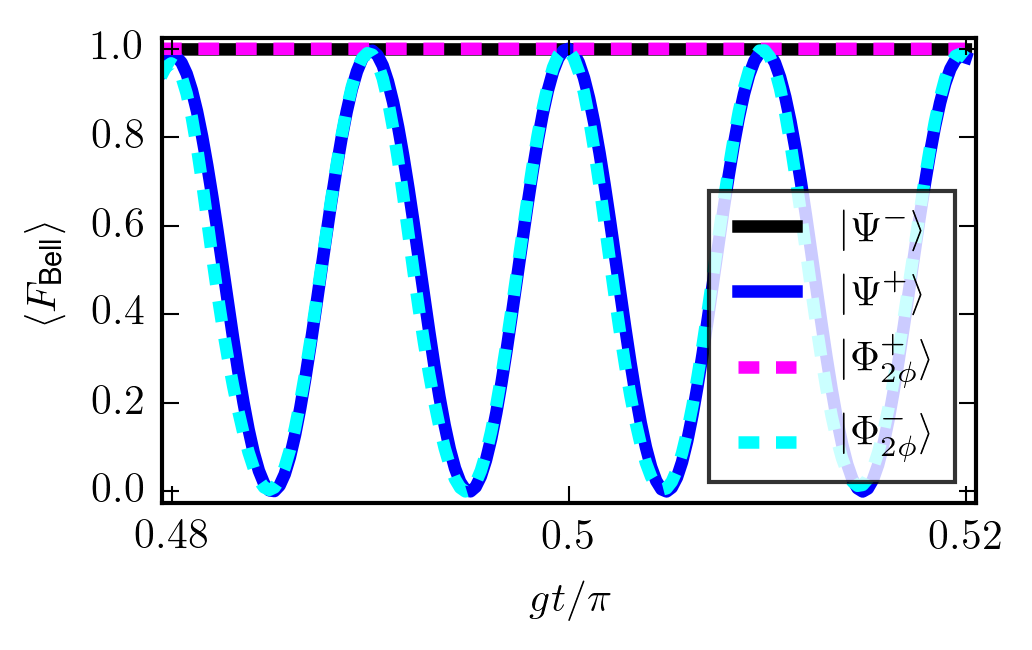}
  \caption{\label{Bellfidtimeig}
 Ensemble average fidelity $\langle F_{\rm Bell}\rangle$ for each postselected Bell state after the application of the protocol around the optimal time $t_{r}/2$.}
\end{figure}

As the protocol is envisioned  to work at half of the revival time, it is important to explore the
sensitivity of the Bell measurement when the interaction time is closed but not exactly $t_{r}/2$.
In Fig.~\ref{Bellfidtimeig} we plot the average fidelity in a short time window close to half of the revival time for  the four
postselected Bell states. In this case we have set the mean photon number $\bar{n}=50$, for which we know almost perfect fidelity can be obtained. The results show almost unit fidelity for projecting onto the states $\ket{\Psi^{-}}$ and $\ket{\Phi^{+}_{2\phi}}$. The first case can be understood in analogy to the Tavis-Cummings model for which one can show that the state $\ket{\Psi^{-}}$ 
remains invariant under time evolution \cite{Torres2014}.  
Approximately regular oscillating behavior is found for the other two Bell states near the optimal time $gt=\pi/2$,
with a similar effective frequency  roughly given by $g(\bar{n}+1)$. 
In order to get an idea of a deviation $\epsilon$ allowed in the interaction time $t_r/2 \pm\epsilon$, one can estimate that this possible error must satisfy the condition $|\epsilon|\ll 1/2\pi g(\bar{n}+1)$
for the protocol to work with nearly optimal fidelity.  

\section{Discussion}
\label{discusion}
We now comment on the possibilities of an experimental realization of the above mentioned protocol. 
In the cavity-QED scenario, experiments using optical conveyor belts to transport neutral laser-cooled cesium atoms into an optical resonator have been successfully performed \cite{Reimann2014,Reimann2014NJP}. 
Adapted to our scheme, the idea would be to transport atoms using a standing wave dipole trap 
into a sequence of two single mode optical cavities. 
The effects of losses can be neglected, provided 
the experiment operates in the strong coupling regime, where  
the condition $g\gg \gamma, \kappa$ holds, i.e., the qubit-field coupling is much larger than the spontaneous decay rate $\gamma$ of the atoms 
and the rate accounting for photon losses in the cavities $\kappa$.
As in our model the revival time is independent of the mean photon number, at the optimal time these
conditions are slightly modified to: $g/\gamma\gg\pi/2$ and $g/\kappa\gg \pi/2$. These requirements can be satisfied in current cavity-QED experiments \cite{Hamsen17}.

Similar schemes involving coupled cavities and two-level atoms  have  been  proposed  in the context of coherence and entanglement protection in the presence of dissipation \cite{LoFranco2015}. The physical realization of these  systems seems to fit very well within the context of the circuit quantum electrodynamics architecture (circiut-QED) \cite{Blais2004,Bronn2015,Nori2017}, 
where transmon qubits can be efficiently coupled to coplanar waveguide cavities. 
High fidelity preparation of 
entangled input initial states for the protocol can be in principle
engineered using a quantum bus trough a transmission line resonator as described in Ref.~\cite{DiCarlo2009}, and the interaction time of the two-qubit system 
with the resonator mode can also be switched-off after the corresponding projective measurement of the optical mode, thus implementing all the steps of the algorithm
in a single on-chip superconducting circuit.
 
It is worth commenting on the possible  implementation in trapped-ion experiments, as they constitute one of the most successful platforms for quantum simulation and quantum information processing  \cite{Leibfried2003}. An interesting simulation based on trapped ions has been recently proposed for emulating the dynamics of the two-photon Rabi model in different coupling regimes \cite{Solano}.  As our protocol is supposed to work in the so called strong-coupling regime, where the qubit frequency is assumed to be very small compared with the qubit-field coupling, its implementation in this particular architecture seems to be  feasible with current technology. In this context two identical ions with two internal electronic states are placed in a harmonic trap of a definite frequency.
By driving the  ions with a laser source at the second red side-band, 
an effective nonlinear two-phonon coupling can be generated \cite{Solano,Plenio}. After the application of a vibrational RWA one can obtain 
the trapped-ion Hamiltonian in the interaction picture:
$\tilde{H}_{I}=-\Omega\eta^2(S_{\rm eg}a^2+S_{\rm ge}{a^{\dagger}}^2)/2$,
which can be immediately recognized as our effective Hamiltonian model
in Eq.~\eqref{interaction2} with an effective coupling given by $g_{\rm{eff}}=-\Omega\eta^{2}/2$, being $\Omega$ the Rabi frequency of the laser, and $\eta$ the so-called Lamb-Dicke parameter.  
The Rabi frequency in trapped-ion experiments lies in the range of $\rm{kHz}$ and $\eta\sim 10^{-2}$, which leads to a coupling constant of $g_{\rm{eff}}\sim 10^{2}$ $\rm{Hz}$. Taking into account these values,  it is possible to estimate the optimal time needed to perform the Bell measurement scheme in a trapped-ion setup from the relation $g_{\rm{eff}}t/\pi=1/2$, which results in a simulation time of $\sim 10 \rm{ms}$. Typical experiments involving, for instance, optical $^{40}$Ca$^+$ ions have coherence times of $\sim$~3ms \cite{Gerritsma2011,Timoney2011}. These times are still small if 
a Bell measurement based on the proposed protocol is to be performed. However, continuous dynamical decoupling schemes have been recently proposed in order to achieve long-time coherent dynamics
by eliminating magnetic dephasing noise in ion-trap simulators \cite{Plenio}, making our proposal for Bell state discrimination more
realistic and in reach of current technology.

 In our treatment we have assumed, for simplicity, perfect projections onto coherent states in order to discriminate the field components. A more realistic discrimination scheme can be achieved by means of a
 balanced homodyne detection (BHD), where the field state is superposed with a strong laser beam, known as local oscillator (LO), in a 50/50 beam splitter (BS). 
By measuring both outputs of the BS via photodetectors, one can achieve the projection 
onto eigenstates $\ket{x,\theta_L}$ of the field quadratures $x_{\theta_L}=(a^\dagger e^{i\theta_L}+a e^{-i\theta_L})/2$, whose eigenvalue $x$ is proportional to the recorded photon difference \cite{Paris}. The selected quadrature depends on the phase $\theta_L$ of the LO. 
In our protocol, the optimal  value is $\theta_L=\phi$ in order to distinguish the field components at time $t_r/2$.  After the BHD, the field is projected and the atoms collapse to the state  $\rho_{\rm at}={\rm Tr}_{f}[\ketbra{x,\phi}{x,\phi}\rho]$,
with the pure state $\rho=\proj{\Psi(t_r/2)}$ in Eq. \eqref{Psiatt2}. With almost unit
probability, the postselected atomic state is one of the pure states resulting from the measurement operators 
$M_\phi^\pm$ in \eqref{projat} applied to the state \eqref{psi}.  This follows from the fact that the field is solely composed of two
coherent states, whose overlaps with the measured field quadrature are given by 
$\braket{x,\phi}{\pm\alpha}=(2/\pi)^{1/4} \exp[-(x\pm|\alpha|)^2]$ \cite{Torres2014}, i.e., probable measurements 
 are only registered for values of $x$ close to $\pm |\alpha|$. Therefore, in a BHD of the cavity field,  positive (negative) photon counting corresponds to the atomic measurement operator $M_\phi^+$ ($M_\phi^-$),  in Eq. \eqref{projat}, of the initial atomic state given by Eq. \eqref{psi}. Imperfect efficiency ($\varepsilon\in[0,1]$) of the photodetectors can also be considered. In this case the POVM describing the BHD is composed of Gaussian convolution of the ideal projectors onto the field quadratures with variance $\Delta_\varepsilon^2=(1-\varepsilon)/4\varepsilon$ \cite{Paris}. The recorded probability distribution for each coherent state will then have an effective variance 
of $\Delta_\varepsilon^2+1/4$. Therefore, one can be confident that BHD is still applicable if $4(\Delta_\varepsilon^2+1/4)^{1/2}$ (four times
the  standard deviation) is smaller than the distance $|\alpha|$ of each coherent state to
 the origin.
This restriction translates into $\varepsilon> 4/|\alpha|^2$ in terms of the photodetection efficiency.

 \section{Conclusions}
 \label{conclusions}
We have presented a Bell measurement scheme on atomic qubits that interact 
with the electromagnetic field contained in two separate cavities via two-photon processes in 
a two-stage Ramsey-type setup. The protocol is based on  the two-atom 
two-photon Dicke model in the limit of large photon number for initial coherent states of the field. Under such conditions, we have derived an approximate solution 
 in terms of atomic Bell states and photonic coherent states, allowing the 
identification of an appropriate Bell-measurement protocol via coherent state discrimination in two separate cavities. In contrast with previous proposed protocols \cite{Torres2014,Torres2016} based on multiphoton states, the one presented here allows a complete discrimination of the four atomic Bell states, i.e. a 100\%-efficient
Bell measurement that we have numerically confirmed  by computing the average fidelity over 
random initial states. The robustness of the protocol as a function  of the interaction time has been tested and the corresponding condition for a possible error in terms of the mean photon number was estimated. 
By analyzing the time-dependent state 
of the full system, we have also demonstrated that tripartite entangled GHZ states 
can be naturally generated by the unitary dynamics of the two-photon model, 
a possibility that can be further exploited in other quantum information protocols. 
It is worth stressing that the complete projection onto the full Bell basis is possible  
as a consequence of the  perfect discrimination of two separate components 
of the evolved field which in turn relies on the perfect
revivals of Rabi oscillations in the two photon model. 

\begin{acknowledgments}
We thank Ralf Betzholz and Thomas H. Seligman for enriching discussions. 
CGG is grateful to CONACYT for financial support under doctoral fellowship  385108 and research grant 219993, and to UNAM/DGAPA/PAPIIT for the research grants IG100616 and IN103017.  JMT acknowledges support by  PRODEP-SEP project 511-6/18-9344. 
\end{acknowledgments}

\end{document}